\documentclass[12pt]{article}
\usepackage{a4wide}
\usepackage{epsfig}
\usepackage{amsmath}

\newlength{\absize}
\catcode`@=11
\def\citer{\@ifnextchar [{\@tempswatrue\@citexr}{\@tempswafalse\@citexr[]}}

\def\@citexr[#1]#2{\if@filesw\immediate
  \write\@auxout{\string\citation{#2}}\fi
  \def\@citea{}\@cite{\@for\@citeb:=#2\do
    {\@citea\def\@citea{--\penalty\@m}\@ifundefined
       {b@\@citeb}{{\bf ?}\@warning
       {Citation `\@citeb' on page \thepage \space undefined}}%
\hbox{\csname b@\@citeb\endcsname}}}{#1}}
\catcode`@=12

\begin{document}
  \thispagestyle{empty}
  \pagestyle{empty}
  \renewcommand{\thefootnote}{\fnsymbol{footnote}}
\newpage\normalsize
    \pagestyle{plain}
    \setlength{\baselineskip}{4ex}\par
    \setcounter{footnote}{0}
    \renewcommand{\thefootnote}{\arabic{footnote}}
\newcommand{\preprint}[1]{%
  \begin{flushright}
    \setlength{\baselineskip}{3ex} #1
  \end{flushright}}
\renewcommand{\title}[1]{%
  \begin{center}
    \LARGE #1
  \end{center}\par}
\renewcommand{\author}[1]{%
  \vspace{2ex}
  {\Large
   \begin{center}
     \setlength{\baselineskip}{3ex} #1 \par
   \end{center}}}
\renewcommand{\thanks}[1]{\footnote{#1}}
%\begin{center}
%{Physical Review Letters (in press)}
%\end{center}
%\begin{flushright}
%
%\end{flushright}
\vskip 0.5cm

\begin{center}
{\large \bf Angular Momentum of Supersymmetric Cold Rydberg Atoms}
\end{center}
\vspace{1cm}
\begin{center}
Jian-zu Zhang
\end{center}
%-----------------------------------
%   Address
%-----------------------------------
\vspace{1cm}
\begin{center}
East China Institute for Theoretical Physics, 130 Mei Long Road,
Shanghai 200237, China and School of Science, East China
University of Science and Technology, Shanghai 200237, China
\end{center}
\begin{center}
(Received 19 October 1995)
\end{center}
\vspace{1cm}
%%%%%%%%%%%%%%%%%%%%%%%%%%%%%%%%%%%%%%%%%%%%%%%%%%%%%%%%%%%%%%

\begin{abstract}
Semiunitary transformation is applied to discuss
supersymmetrization of cold Rydberg atoms. In the limit of
vanishing kinetic energy the lowest angular momentum of the
supersymmetric cold Rydberg atom is $3\hbar/2$. A possible
experimental verification is suggested. [S0031-9007(96)00109-3]

\vspace{0.3cm}


\end{abstract}

\clearpage
%%%%%%%%%%%%%%%%%%%%%%%%%%%%%%%%%%%%%%%%%%%%%%%%%%%%%%%%%%%%%%%%%%%%%
Recently Baxter \cite{Baxt} showed that cold Rydberg atoms can
play an interesting role of realizable analogs of Chern- Simons
theory \cite{DJT82,DJT90}. By choosing an atomic dipole of a cold
Rydberg atom in electric and magnetic field, and by an appropriate
experimental arrangement the motion of the dipole is constrained
to be planar and rotationally symmetric, the R\"otgen interaction
takes on the form of a Chern-Simons term \cite{DJT90}. By placing
the dipole in a strong magnetic field and in an appropriate
optical trapping field, the elimination of the kinetic energy term
in the Lagrangian could be achieved physically and Baxter
\cite{Baxt} showed that the canonical angular momentum spectrum
changes from one consisting of integers to one consisting of
positive half integers; thus in principle an experimental
verification of the Chern-Simons feature of fractional angular
momentum is allowed.

Reference \cite{KN} showed evidence for a phenomenological
supersymmetry in atomic physics. In $SS$ $QM$ the term
¡°supersymmetry¡± has nothing to do with spins at all, its meaning
is just as that $SS$ $QM$ is represented by a pair of bosonic
Hamiltonians $H_{-}$ and $H_{+}$ which are superpartners of the
$SS$ Hamiltonian $H_{s}=H_{-}+H_{+}$, and SS charge $Q_s$,
$Q_s^{\dagger}$, and $H_{s}$ satisfy $SS$ algebra \cite{Witt}.

In $SS$ $QM$ the form of the problem must be truly one
dimensional. For the one-dimensional system there is no room to
define the angular momentum; one cannot discuss the relation of
spectra of angular momenta between the system and its
superpartner. For the case of two-dimensional Rydberg atoms
angular momentum can be defined as an exterior product. This
system can be reduced into two uncoupled one-dimensional systems,
thus the supersymmetrization of its Hamiltonian and angular
momentum can be simultaneously performed by a semiunitary
transformation ($SUT$) which will be defined later. This for the
first time allows in principle a verification of the relation of
spectra of angular momentum between a system and its superpartner.
The results show that in the limit of vanishing kinetic energy the
angular momentum spectrum of the $SS$ cold Rydberg atom changes
from one consisting of integers to one consisting of positive half
integers, with the lowest angular momentum $3\hbar/2$; $SUT$
destroys the lowest angular momentum state with $\hbar/2$ of the
cold Rydberg atom.

{\it Cold Rydberg atom}.---The Lagrangian of the cold Rydberg atom
with an atomic dipole $d$ in electric and magnetic field is
\cite{Baxt}
%%%
\begin{equation}
\label{Eq:L}%1e
L=\frac{1}{2}M\dot{R}_i\dot{R}_i+\frac{1}{2}g\epsilon_{ij}R_i\dot{R}_j
-\frac{1}{2}KR_i R_i.
\end{equation}
%%%
where $M$ is the mass of the dipole, the electric field $\vec{E}$
acts radially in the $x-y$ plane, $E_i \sim -R_i \;(i= 1,2)$, and
the constant magnetic field $\vec{B}$ aligns the $z$ axis. The
constant parameters $g$ and $K$ are proportional to the magnitude
of the dipole moment and, respectively, magnetic and electric
field dependent. $\epsilon_{12}=-\epsilon_{21}=1;$
$\epsilon_{11}=\epsilon_{22}=0.$
In (\ref{Eq:L}) the second term
$\frac{1}{2}g\epsilon_{ij}R_i\dot{R}_j$
takes on the form of the Chern-Simons interaction. The Hamiltonian
obtained from (\ref{Eq:L}) is
%%%
\begin{eqnarray}
\label{Eq:CS-H1}%2e
H=\frac{1}{2M}\left( p_i+\frac{g}{2}\epsilon_{ij} R_j\right)^2
+\frac{K}{2}R_iR_i
=\frac{1}{2M}P_i P_i+\frac{1}{2M}g\epsilon_{ij} P_i
R_j+\frac{1}{2}M\Omega^2 R_iR_i,
\end{eqnarray}
%%%
where the canonical momentum
$P_i=M\dot{R}_i-g\epsilon_{ij}R_j/2$,
and the frequency
%%%
\begin{equation}
\label{Eq:Omega}%3e
\Omega=\left(g^2/4M^2+K/M\right)^{1/2}.
\end{equation}
%%%
In (\ref{Eq:Omega}) the dispersive ¡°mass¡± term $g/2M$ comes from
the presence of the Chern-Simons term. By changing the variables
$R_i$ , $P_i$ to
%%%
\begin{subequations}
\begin{eqnarray}
\label{Eq:Xa}%4ae
X_a&=&(M\Omega/2\omega_a)^{1/2}R_1-(1/2M\Omega\omega_a)^{1/2}P_2,
\nonumber\\
X_b&=&(M\Omega/2\omega_b)^{1/2}R_1+(1/2M\Omega\omega_b)^{1/2}P_2,
\end{eqnarray}
\begin{eqnarray}
\label{Eq:Pa}%4be
P_a&=&(\omega_a/2M\Omega)^{1/2}P_1+(M\Omega\omega_a/2)^{1/2}R_2,
\nonumber\\
P_b&=&(\omega_b/2M\Omega)^{1/2}P_1-(M\Omega\omega_b/2)^{1/2}R_2
\end{eqnarray}
\end{subequations}
%%%
where
%%%
\begin{equation}
\label{Eq:omega-ab}%5e
\omega_a = \Omega + g/2M, \; \omega_b = \Omega - g/2M,
\end{equation}
%%%
the Hamiltonian is rewritten in the form of two uncoupled harmonic
oscillators of unit mass and of frequencies $\omega_a$ and
$\omega_b$. Define
%%%
\begin{eqnarray}
\label{Eq:Aab}%6e
A_\alpha &=& \left(\hbar/2\omega_\alpha\right)^{1/2} d/dX_\alpha +
\left(\omega_\alpha/2\hbar\right)^{1/2} X_\alpha,
\nonumber \\
A^{\dagger}_\alpha &=& - \left(\hbar/2\omega_\alpha\right)^{1/2}
d/dX_\alpha + \left(\omega_\alpha/2\hbar\right)^{1/2} X_\alpha
\quad (\alpha = a,b);
\end{eqnarray}
%%%
the Hamiltonian (\ref{Eq:CS-H1}) becomes
%%%
\begin{equation}
\label{Eq:H-ab}%7e
H = H_a + H_b, \quad
H_{\alpha} = \hbar\omega_{\alpha} \left(A^{\dagger}_{\alpha}
A_{\alpha} + \frac{1}{2}\right) \quad (\alpha = a,b).
\end{equation}
%%%
If $|n_{\alpha}\rangle$ is an eigenfunction of $H_{\alpha}$ with
eigenvalue
$E_{n_{\alpha}}=\hbar\omega_{\alpha}(n_{\alpha}+1/2)\quad (\alpha
= a,b),$
$|n_{a},n_{b}\rangle=|n_{a}\rangle|n_{b}\rangle$
is an eigenfunction of H with eigenvalue
%%%
\begin{equation}
\label{Eq:E-ab}%8e
E_{n_{a},n_{b}}=\hbar\omega_{a}(n_{a}+\frac{1}{2})+
\hbar\omega_{b}(n_{b}+\frac{1}{2}) \quad
(n_{a},n_{b} =0,1,2,\dots),
\end{equation}
%%%

In two dimensions the canonical angular momentum is defined as an
exterior product which is a scalar
%%%
\begin{equation}
\label{Eq:J}%9e
J = \epsilon_{ij}R_iP_j = J_a-J_b, \quad
J_\alpha = \hbar(A^{\dagger}_\alpha A_\alpha + \frac12) \quad
(\alpha = a,b).
\end{equation}
%%%
Because $[H,J] = 0$, $J$ is a constant of motion, and $J$ and $H$
have common eigenfunctions $|n_{a},n_{b}\rangle$. Notice that
$J_a$ and $J_b$ have zero-point angular momentum $\hbar/2$.
Because of the cancellation of $\hbar/2$ between modes $a$ and
$b$, the eigenvalues of $J$ are integer multiples of $\hbar$,
%%%
\begin{equation}
\label{Eq:J-ab}%10e
j_{n_a,n_b} = \hbar(n_a-n_b) \quad (n_a,n_b = 0,1,2, \dots).
\end{equation}
%%%

%%%%%%%%%%%%%%%%%%%%%%%%%%%%%%%%%%%%%%%%%%%%%%%%%%%%%%%%%%%%%%%%%
{\it SS QM}. -- In one-dimensional $SS$ $QM$ a quantum system is
described by a pair of related bosonic Hamiltonians $H_{-}$ and
$H_{+}$ \cite{Witt}:
%%%
\begin{equation}
\label{Eq:H-x}%11e
H_{\pm} = - \frac{\hbar^2}{2} \frac{d^2}{dx^2 } + V_{\pm}(x).
\end{equation}
%%%
Suppose the ground-state wave function
$\psi^{(-)}_0(x) \equiv \psi_0(x)$
of $H^{(-)}$ of $H_{-}$ is known, consider only unbroken
supersymmetry corresponding to a normalizable ground state
$\psi_0$ and adjust the ground-state energy $E^{(-)}_0 = 0$.
Introducing the superpotential
$W(x) = -(\hbar/\sqrt{2})\psi^\prime_0/\psi_0$
(a prime denotes $d/dx$), we have
$V_{\pm}(x) = W^2(x) \pm (1/\sqrt2)W^\prime(x).$
The Hamiltonians $H_{\pm}$ can be factorized as
%%%
\begin{equation}
\label{Eq:H-A}%12e
H_{-}= A^{\dagger}A, \quad H_{+} = AA^{\dagger},
\end{equation}
%%%
where
%%%
\begin{equation}
\label{Eq:A-A}%13e
A = \frac{\hbar}{\sqrt{2}}\frac{d}{dx} + W(x), \quad
A^{\dagger} = - \frac{\hbar}{\sqrt2} \frac{d}{dx} + W(x), \quad
[A,A^{\dagger}] = \sqrt2 \hbar W^\prime(x).
\end{equation}
%%%
The eigenfunctions $\psi^{-}_n$ and $\psi^{+}_n$ of $H_{-}$ and
$H_{+}$, respectively, with eigenvalues $E^{-}_n$ and $E^{+}_n$
are related by
\begin{subequations}
\begin{eqnarray}
\label{Eq:E+-a}%14ae
E^{(+)}_n = E^{(-)}_{n+1},
\end{eqnarray}
\begin{eqnarray}
\label{Eq:E+-b}%14be
\psi^{(+)}_n = \left[E^{(-)}_{n+1} \right]^{-1/2}A\psi^{(-)}_{n+1}
\quad (n = 0,1,2, \dots).
\end{eqnarray}
\end{subequations}
%%%
We find that the superpartner Hamiltonians  $H_{-}$ and $H_{+}$
can be related by a $SUT$ \cite{ZLN}.
%[6].
Consider the operator $AA^{\dagger}$. Because
$AA^{\dagger}\psi^{(+)}_n=E^{(+)}_n\psi^{(+)}_n, \quad
E^{(+)}_n>0 \quad (n = 0,1,2, \dots),$
$AA^{\dagger}$ is positive definite, we can define
%%%
\begin{equation}
\label{Eq:U}%15e
U = \left( AA^{\dagger}\right)^{-1/2} A, \quad
U^{\dagger} = A^{\dagger} \left(AA^{\dagger}\right)^{-1/2},
\end{equation}
%%%
$U$ and $U^{\dagger}$ satisfy
%%%
\begin{equation}
\label{Eq:U-U}%16e
UU^{\dagger} = I, \quad
U^{\dagger}U = A^{\dagger}\left(AA^{\dagger}\right)^{-1}A \equiv
Q.
\end{equation}
%%%
Notice that
$A^{\dagger}A\psi^{(-)}_n = E^{(-)}_n\psi^{(-)}_n$,
$E^{(-)}_0 = 0, \quad E^{(-)}_n > 0$
for $n \ge 1$, so $A^{\dagger}A$ is semipositive definite and
$(A^{\dagger}A)^{-1}$ is singular. Thus we cannot use the operator
identities
$f(AA^{\dagger})A = Af(A^{\dagger}A)$,
and
$A^{\dagger}f(AA^{\dagger}) = f(A ^{\dagger}A)A^{\dagger}$;
the operator Q defined in (\ref{Eq:U-U}) is not a unit operator.
The operator $U$ defined in (\ref{Eq:U}) which satisfies
(\ref{Eq:U-U}) is called semiunitary \cite{note-1}. Under this
$SUT\quad$ $H_{-}$ is transformed into $H_{+}$,
%%%
\begin{equation}
\label{Eq:H-U}%17e
H_{+}=UH_{-}U^{\dagger}.
\end{equation}
%%%
Correspondingly, $\psi^{(-)}_n$ is transformed into
$U\psi^{(-)}_n$. Notice that $U\psi_0=0$,  \quad
$U\psi^{(-)}_{n+1} = [ E^{(-)}_{n+1}]^{-1/2} A\psi^{(-)}_{n+1}
=\psi^{(+)}_n,$
with eigenvalues
$E^{(+)}_n = E^{(-)}_{n+1} \quad  (n = 0,1,2, \dots)$.
The ground energy $E^{(+)}_0 = E^{(-)}_1>0$; here $E^{(-)}_0 = 0$
disappears in the spectrum of $H_{+}$. Thus the $SUT$ transforms
the $(-)$ system into its superpartner system $(+)$, which fully
covers the results of $SS$ $QM$.

The operator $Q$ defined in (\ref{Eq:U-U}) satisfies
%%%
\begin{equation}
\label{Eq:Q}%18e
Q^2 = Q, \quad [Q,H_{-}] = 0.
\end{equation}
%%%
$Q$ is a projection operator, and its eigenvalues $q \quad (q = 0,
1)$ are good quantum numbers. $Q$ and $H_{-}$ have common
eigenfunctions. Because
$Q\psi_0 = 0, \quad Q\psi^{(-)}_{n+1} = \psi^{(-)}_{n+1}$,
we denote
$|\psi_0 \rangle = |0,0 \rangle, \quad
|\psi^{(-)}_{n+1} \rangle = |n+1,1 \rangle \quad (n=0,1,2,\dots)$,
which are common eigenstates of  $H_{-}$ and $Q$. The Hilbert
space $\cal H$ is divided into two subspaces ${\cal H}_{0}$ and
${\cal H}_{1}$, consisting of eigenstates
$|0,0 \rangle$
and
$|n+1,1 \rangle$
with, respectively, eigenvalues $q = 0$ and $1$, $\quad$
${\cal H}= {\cal H}_0 \oplus {\cal H}_1$.
In the basis $|0,0 \rangle$ and $|n+1,1 \rangle$ $\quad$
$Q = I- |0,0 \rangle \langle 0,0|$,
thus in the ${\cal H}_{1}$ subspace the operator $Q$ reduces to a
unit operator and the $SUT$ reduces to a unitary transformation.

Now we compare a unitary transformation and a $SUT$. In quantum
mechanics a unitary operator maintains all the physical properties
of a quantum system. The situation is different for $SUT$. The
main reason is that $SUT$ is a singular operator which readily
appears in the structure of the projector $Q$ defined in
(\ref{Eq:U-U}). Because of such singularities, $SUT$ only partly
maintains physical properties of the $(-)$ system. For example, it
maintains the complete relation of $\psi^{(-)}_n$, but does not
maintain the orthogonality and normalization of $\psi^{(-)}_n$,
and only partly maintains the eigenvalues of $H_{-}$. However, in
the subspace ${\cal H}_{1}$ the $SUT$ behaves just like a unitary
transformation. $SUT$ transforms the $(-)$ system defined in the
full Hilbert space ${\cal H}$ into the $(+)$ system defined in the
subspace ${\cal H}_{1}$, destroys the ground state of $H_{-}$, but
maintains all other physical properties of the $(-)$ system in the
subspace ${\cal H}_{1}$. This explains the legitimacy of the
applications of $SUT$ to $SS$ $QM$.

{\it Supersymmetrization of cold Rydberg atoms}. -- Using $SUT$ we
can thoroughly study the supersymmetrization of cold Rydberg
atoms. The Hamiltonian (\ref{Eq:H-ab}) has two modes $a$ and $b$.
We define the potential
$V^{(-)}(X) = \sum_{\alpha=a,b} (\frac{1}{2}\omega^2_\alpha
X^2_\alpha - \frac{1}{2}\hbar\omega_\alpha)$,
with the Hamiltonian $H_{-}$ and the angular momentum $J_{-}$
represented by $A_\alpha$ and $A^\dagger_\alpha$ defined in
(\ref{Eq:Aab}),
%%%
\begin{equation}
\label{Eq:H-AA}%19e
H_{-} = H_{-a} + H_{-b}, \quad
H_{-a} = \hbar\omega_a A^\dagger_a A_a \otimes I_b, \quad
H_{-b} = \hbar\omega_b I_a \otimes A^\dagger_b A_b.
\end{equation}
\begin{equation}
\label{Eq:J-AA}%20e
J_{-} = J_{-a} - J_{-b}, \quad
J_{-a} = \hbar\left(A^\dagger_a A_a + \frac12 \right) \otimes I_b,
\quad
J_{-b} = \hbar I_a \otimes \left(A^\dagger_b A_b + \frac12
\right).
\end{equation}
%%%
where $I_a$ and $I_b$ are unit operators in the Hilbert spaces
${\cal H}_a$ and ${\cal H}_b$, respectively, corresponding to
modes $a$ and $b$. The common eigenstates of $H_{-}$ and $J_{-}$
are
$|n_a,n_b \rangle_{-} = |n_a\rangle |n_b \rangle$
with, respectively, eigenvalues
%%%
\begin{eqnarray}
\label{Eq:E-ab}%21e
E^{(-)}_{n_a,n_b} &=& \hbar\omega_a n_a + \hbar\omega_b n_b,
\nonumber\\
j^{(-)}_{n_a,n_b} &=& \hbar(n_a + \frac12) - \hbar(n_b + \frac12)
\nonumber\\
(n_a,n_b &=& 0,1,2,\dots)
\end{eqnarray}
%%%
$SUT$ is defined as
%%%
\begin{eqnarray}
\label{Eq:U-ab}%22e
U &=& U_a \otimes U_b, \quad U^\dagger = U^\dagger_a \otimes
U^\dagger_b, \quad
\nonumber\\
U_\alpha &=& (A_\alpha A^\dagger_\alpha)^{-1/2}A_\alpha, \quad
U^\dagger_\alpha = A^\dagger_\alpha (A_\alpha
A^\dagger_\alpha)^{-1/2} \quad
%\nonumber\\
%
(\alpha = a,b)
\end{eqnarray}
%%%
with
%%%
\begin{equation}
\label{Eq:U-U-2}%23e
UU^\dagger = I, \quad U^\dagger U = Q,
\end{equation}
%%%
where
\begin{eqnarray}
\label{Eq:Q-ab}%24e
Q = Q_a \otimes Q_b, \quad
Q_\alpha = A^\dagger_\alpha (A_\alpha
A^\dagger_\alpha)^{-1}A_\alpha \ne I_\alpha \quad
%\nonumber\\
%
(\alpha = a,b)
\end{eqnarray}
%%%
Under $SUT$ (\ref{Eq:U-ab}), $H_{-}$ and $J_{-}$ are transformed
into
%%%
\begin{eqnarray}
\label{Eq:H-AAab}%25e
H_{+} &=& UH_{-} U^\dagger = H_{+a} + H_{+b},
\nonumber\\
H_{+a} &=& \hbar\omega_a A_a A^\dagger_a \otimes I_b, \quad
H_{+b} = \hbar\omega_b I_a \otimes A_b A^\dagger_b.
\end{eqnarray}
%%%
\begin{eqnarray}
\label{Eq:J-AAab}%26e
J_{+} &=& UJ_{-} U^\dagger = J_{+a} - J_{+b},
\nonumber\\
J_{+a} &=& \hbar (A_a A^\dagger_a \otimes I_b + \frac{1}{2} I_a
\otimes I_b), \quad
J_{+b} = \hbar (I_a \otimes A_b A^\dagger_b + \frac{1}{2} I_a
\otimes I_b).
\end{eqnarray}
%%%
Correspondingly, under $SUT$ (\ref{Eq:U-ab}) $|n_a,n_b
\rangle_{-}$ is transformed into $U|n_a,n_b \rangle_{-}$. Notice
that
$U|0,0 \rangle_{-}=0$, $U|0,n_b \rangle_{-}=0$ and $U|n_a,0
\rangle_{-}=0$
which shows that the states $|0,0 \rangle_{-}$, $|0,n_b
\rangle_{-}$, and $|n_a,0 \rangle_{-}$ are destroyed by $SUT$.
Thus
$|n_a,n_b\rangle_{+} =  U|n_a + 1, n_b +1 \rangle_{-}
= [ E^{(-)}_{n_a+1} E^{(-)}_{n_b+1}]^{-1/2} A_a|n_a +
1 \rangle_{-} \otimes A_b|n_b+1\rangle_{-} \quad
(n_a,n_b = 0,1,2,\dots).$
The common eigenstates of $H_{+}$ and $J_{+}$ are
$|n_a,n_b\rangle_{+} =  U|n_a + 1, n_b +1 \rangle_{-}$
with, respectively, eigenvalues
%%%
\begin{eqnarray}
\label{Eq:E-ab+}%27e
E^{(+)}_{n_a,n_b} &=& \hbar\omega_a (n_a + 1) +
\hbar\omega_b (n_b + 1),
\nonumber\\
j^{(+)}_{n_a,n_b} &=& \hbar(n_a + \frac32) - \hbar(n_b + \frac32)
\nonumber\\
(n_a,n_b &=& 0,1,2,\dots)
\end{eqnarray}
%%%
For the $(+)$ system the ground state $|0,0\rangle_{+}$ has energy
$E^{(+)}_{0,0} = \hbar(\omega_a + \omega_b) = 2\hbar\Omega$.
Here the states
$|0,0\rangle_{-}$, $|0,n_b\rangle_{-}$, and $|n_a,0\rangle_{-}$
of $H_{-}$ with energy
$E^{(-)}_{0,0} = 0$, $E^{(-)}_{0,n_b} = \hbar\omega_b n_b$, and
$E^{(-)}_{n_a,0} = \hbar\omega_a n_a$ disappear.

Because of the cancellation of the angular momenta between modes
$a$ and $b$ the angular momenta $J_{-}$ and $J_{+}$ have the same
spectrum
$j^{(-)}_{n_a,n_b} = \hbar(n_a - n_b) = j^{(+)}_{n_a,n_b} =$ any
integer multiple of $\hbar$.
The $(+)$ system still has the same lowest eigenvalue as the $(-)$
system
$j^{(+)}_{0,0} = j^{(-)}_{0,0} = 0$.

Now we consider the interesting limit case $M \to 0$ discussed by
Baxter \cite{Baxt}. In this case there are constraints which
should be carefully considered. The first equation of
(\ref{Eq:CS-H1}) shows that the limit case $M \to 0$ requires the
constraints
%%%
\begin{equation}
\label{Eq:Ci}%28e
C_i=P_i+g\epsilon_{ij} R_j/2=0.
\end{equation}
%%%
We observe that the Poisson brackets
$\{C_i,C_j\}=g\epsilon_{ij}\ne 0$
\cite{note-2}, so that the Dirac brackets can be determined
\cite{MZ}
%%%
\begin{equation}
\label{Eq:Dirac}%29e
\{R_1,P_1\}_D=\{R_2,P_2\}_D = 1/2,\; \{R_1,R_2\}_D = -1/g, \;
\{P_1,P_2\}_D = -g/4.
\end{equation}
%%%
Other Dirac brackets of $R_i$ and $P_i$ are zero. The Dirac
brackets of $C_i$ with any variables $R_i$ and $P_i$ are zero;
thus (\ref{Eq:Ci}) are strong conditions in the sense of Dirac
which can be used to eliminate the dependent variables. Choosing
the independent variables $(R_1,P_1)$, (\ref{Eq:Ci}) fixes the
dependent variables
$R_2 = -2P_1/g,\;P_2=gR_1/2.$
In the reduced phase space of the independent variables
$(R_1,P_1)$ the Hamiltonian (\ref{Eq:CS-H1}) has the structure of
a one-dimensional harmonic oscillator
%%%
\begin{equation}
\label{Eq:H1a}%30e
H=\frac{2K}{g^2}P_1^2+\frac{1}{2}K R_1^2
\end{equation}
%%%
According to (\ref{Eq:Dirac}) the quantization condition of the
independent variables is
$[R_1,P_1] = i\hbar/2.$
Set
$R_1=q/\sqrt{2}, \; P_1=p/\sqrt{2}$
which leads to $[q,p] = i\hbar$. Introduce
$V_{-}(q)=\frac{1}{2} m^{\ast}\omega^2 q^2
-\frac{1}{2}\hbar\omega$
where the effective mass $m^{\ast}$ and the frequency $\omega$ are
%%%
\begin{equation}
\label{Eq:m-0mega}%31e
m^{\ast} = g^2/2K, \quad \omega = K/g.
\end{equation}
%%%
Define
%%%
\begin{eqnarray}
\label{Eq:A-A2}%32e
A &=& (\hbar/2 m^{\ast}\omega)^{1/2} d/dq +
(m^{\ast}\omega/2\hbar)^{1/2} q,
\nonumber \\
A^\dagger &=& -(\hbar/2 m^{\ast}\omega)^{1/2} d/dq +
(m^{\ast}\omega/2\hbar)^{1/2} q,
\end{eqnarray}
%%%
we obtain from (\ref{Eq:H1a}
%%%
\begin{equation}
\label{Eq:H1b}%33e
H_{(-)} = \frac{1}{2m^{\ast}}[p^2 + ({m^{\ast}}^2\omega^2 q^2 -
m^{\ast}\hbar\omega)] =
\hbar\omega A^\dagger A
\end{equation}
%%%
with eigenvalues
%%%
\begin{equation}
\label{Eq:E-1}%34e
E^{(-)}_n = n\hbar\omega \quad (n = 0,1,2,\dots).
\end{equation}
%%%
Observe that if we rewrite (\ref{Eq:H1a}) as
$H=P_1^2/2m_1 + m_1\omega_1^2 R_1^2/2$
where the effective mass $m_1 = g^2/4K$ and the frequency
$\omega_1 = 2K/g$, we find that the frequency $\omega$ of
(\ref{Eq:m-0mega}) differs from that of the conventional harmonic
oscillator by a factor of $1/2$. This is the representation of the
well-known fact that reduction to the reduced phase space alters
the symplectic structure \cite{MZ}.

In the limit $M \to 0$ the angular momentum is
%%%
\begin{equation}
\label{Eq:J-1a}%35e
J_{-} = gR_iR_i/2 = \hbar(A^\dagger A + \frac12).
\end{equation}
%%%
The spectrum of $J_{-}$ is a positive half-integer multiple of
$\hbar$,
%%%
\begin{equation}
\label{Eq:j-n}%36e
j^{(-)}_n = \hbar(n + \frac12) \quad (n = 0,1,2,\dots).
\end{equation}
%%%
Comparing (\ref{Eq:H1b}), (\ref{Eq:J-1a}) with (\ref{Eq:H-AA}),
(\ref{Eq:J-AA}), we see that in the limit $M \to 0$  the mode $b$
disappears, only the mode $a$ is maintained \cite{Baxt}. Using $A$
and $A^\dagger$ defined in (\ref{Eq:A-A2}) we construct
$U = (A A^\dagger)^{-1/2}A$
which is a $SUT$. With such a $SUT$, \quad $H_{-}$ in
(\ref{Eq:H1b}) and $J_{-}$ in (\ref{Eq:J-1a}) are simultaneously
transformed into their superpartners
%%%
\begin{eqnarray}
\label{Eq:H-J-2}%37e
H_{+} &=& UH_{-}U^\dagger = \hbar\omega A A^\dagger,
\nonumber\\
J_{+} &=& UJ_{-}U^\dagger = \hbar (A A^\dagger + \frac12),
\end{eqnarray}
%%%
with, respectively, spectra
%%%
\begin{eqnarray}
\label{Eq:E-j-2}%38e
E^{(+)}_n = \hbar\omega (n + 1), \quad
j^{(+)}_n = \hbar (n + 3/2) \quad
(n = 0,1,2,\dots).
\end{eqnarray}
%%%
Comparing with the spectrum of $J_{-}$, the spectrum of $J_{+}$ is
also a positive half integer of $\hbar$, but starting from
$3\hbar/2$, the lowest angular momentum $\hbar/2$ of $J_{-}$ is
destroyed by $SUT$.

A possible experimental verification of the lowest angular
momentum of $SS$ cold Rydberg atoms is allowed in principle. In
the present case the difference between $V_{+}$ and $V_{-}$ is a
constant,
$V_{+} - V_{-} = 2\hbar(g/4M^2 + K/M)^{1/2}$.
Taking the limit of vanishing kinetic energy is achieved as
follows \cite{Baxt}. If the magnetic field is strong enough, the
second term in (\ref{Eq:CS-H1}) is dominant. Further, in an
appropriate optical trapping field the speed of the atom can be
slowed to the extent that the kinetic energy term in
(\ref{Eq:CS-H1}) may be removed \cite{SST}, which leads to the
limit $M \to 0$ \cite{note-3}. Assuming that the planar, confined
dipole is prepared in its energy ground state and interacts with a
radiation mode of a Laguerre-Gaussian form (since a
Laguerre-Gaussian beam carries orbital angular momentum along its
direction of propagation \cite{ABSW}). The expectation value of
the angular momentum in the long time limit shows two distinct
resonances at $\omega_a,\omega_b$. As a diminution in the kinetic
energy term, the $b$ resonance occurs at even greater values of
frequency, until only the $a$ resonance remains achievable at
$\omega_a = K/g$. The supersymmetrization of the $(-)$ system can
be achieved with a constant shift of the potential $ V_{-}$. Thus
the location and nature of possible angular momentum resonances
allow in principle the experimental verification of the lowest
angular momentum spectrum of $SS$ cold Rydberg atoms.

To summarize, the main results obtained in this paper are as
follows: (1) $SUT$ is applied to discuss supersymmetrization of
cold Rydberg atoms. The Hamiltonian and angular momentum of the
$SS$ Rydberg atom are simultaneously obtained. This allows the
possibility of comparing the relation of angular momenta between
the $(-)$ system and $(+)$ system. It is interesting to exploit
further possible applications of $SUT$ in physics. (2) In the
limit of vanishing kinetic energy the lowest angular momentum of
the $SS$ cold Rydberg atom is $3\hbar/2$. The suggested possible
experimental verification provides a crucial test of the idea of
$SS$ $QM$.

\vspace{0.4cm}

This project was supported by the National Natural Science
Foundation of China.

%\clearpage


\begin{thebibliography}{99}
\bibitem{note-1}%7r[7]
The semiunitary operator can exist only in infinite dimension
space. In finite dimension space, if $UU^\dagger = I$, then
$U^\dagger U$ must be a unit operator too. So in finite dimension
space there are only unitary operators. It is easy to show that
semiunitary operators constitute a semigroup.
\bibitem{note-2}%8r[8]
Because of the Poisson brackets $\{C_i,C_j\}=g\epsilon_{ij}\ne 0$,
the consistency requirement $\dot{C_i} = 0$ fixes the Lagrange
multiplier $\lambda_l$ in the first-order Lagrangian with
constraints (\ref{Eq:Ci}). Therefore there is no need to introduce
gauge fixing conditions.
\bibitem{note-3}%11r[11]
The identification of the $M \to 0$ limit with the vanishing
kinetic energy limit requires careful analysis. In the $M \to 0$
limit the Lagrangian $L$ in (\ref{Eq:L}) is reduced to
%
$L_0 = \frac{1}{2}g\epsilon_{ij}R_i\dot{R}_j - \frac{1}{2}KR_i
R_i$,
%
the corresponding canonical momentum is
%
$P_{0i} = \partial L_0/\partial \dot{R_i} =
\frac{1}{2}g\epsilon_{ji}R_j,$
%aaa
and the Hamiltonian is
%
$H_0=P_{0i}\dot{R_i} - L_0 = \frac{1}{2}K R_i R_i.$
%
On the other hand, the Hamiltonian $H$ in (\ref{Eq:CS-H1}) can be
rewritten as
%
$H = \frac{1}{2}M\dot{R_i}\dot{R_i} + \frac{1}{2}KR_i R_i$,
%
where the first term $\frac{1}{2}M\dot{R_i}\dot{R_i}$ is the
kinetic energy term. Thus the kinetic energy $\to 0$ limit gives
%
$H \to H_0 = \frac{1}{2}K R_i R_i$,
%
which justifies the identification of the two limiting processes.
Here the point is that we should clarify the difference between
the kinematical momentum and the canonical momentum whenever
velocity-dependent forces occur.
\end{thebibliography}
\end{document}